\documentstyle[12pt]{article}

\renewcommand{\d}{\delta}
\newcommand{\g}{\gamma}

\newcommand{\e}{\epsilon}

\newcommand{\ar}{\longrightarrow}
\newcommand{\n}{\smallskip}
\newcommand{\nn}{\medskip}

\newcommand{\s}{\sigma}

\newcommand{\la}{\lambda}

\begin{document}
\title{Fast Quantum verification for the formulas of predicate calculus}
\author{Yuri Ozhigov
\thanks{Department of applied mathematics, Moscow state technological
 University "Stankin", Vadkovsky per. 3a, 101472, Moscow, Russia,
 e-mail: \ y@ oz.msk.ru}}
\date{}
\maketitle
\begin{abstract}
Quantum algorithm is constructed which verifies the formulas
of predicate calculus in time $O(\sqrt N )$ with bounded error probability,
where $N$ is the time required for classical algorithms. This 
algorithm uses the polynomial number of simultaneous oracle queries. 
This is a modification of the result of Buhrman, Cleve and Wigderson 9802040.

\end{abstract}

\section{Introduction and Background}

In 1996 L.Grover showed how quantum computer can find the unique solution of equation
$f(x)=1$ in time $O(\sqrt N )$ for a Boolean function $f$ determined by oracle
where $N$ is the number of all possible values for $x$ where classical
algorithm takes the time $\Omega (N)$.

Soon after that M. Boyer, G. Brassard, P. Hoyer and A. Tapp extended this result
to the case of unknown number of solutions. Thus in fact they obtained
the method of verification of a formula $\exists x P(x)$ where $x\in\{ 0,1\}^n $,
$P$ is a predicate, determined by an oracle which instantly returns "$P(x)$ true"
or "$P(x)$ false" for a given $x$. We shall denote
this algorithm by G-BBHT.

The natural generalization of G-BBHT would be the quantum verification of arbitrary 
formula of predicate calculus of the form
\begin{equation}
\forall x_1 \exists y_1 \forall x_2 \exists y_2 \ldots
\exists y_k \ p(x_1 ,y_2 ,\ldots ,x_k ,y_k )
\label{1}
\end{equation}
(prenex normal form).
This is the aim of this article.

Let $N=2^n$ where a string $x_1 y_1 \ldots x_k y_k$ belongs to $\{ 0,1\}^n$.

\newtheorem{Theorem}{Theorem}
\begin{Theorem}
There exists the quantum algorithm which verifies formulas of the form (1)
in time $O(\sqrt N )$ with bounded error probability using $(Cn)^k$ simultaneous 
oracle queries
where constant $C$ depends on the error probability.
\end{Theorem}

Note that in the paper \cite{BCW} H.Buhrman, R.Cleve and A.Wigderson proved
that a formula of the form (1) can be verified in time of order $\sqrt{N} (\log N)^{k-1}$
with only one query at a time (Theorem 1.15). So we can see that the admission of simultaneous
queries gives the corresponding reduction of the time complexity for this problem.
 
Our Theorem is a nontrivial generalization of G-BBHT. The point is that
having no information about $P$, to overcome the change of quantors:
$\forall x\exists y$ we evidently must use a pure classical operation
on G-BBHT: use it as a subroutine in the sequential implementation of
some program for the different cases. The classical realization of such 
approach immediately gives the time $O(\sqrt{|y_k |} |x_1 | |y_1 |\ldots |x_k |)$
which is substantially more than $\sqrt N$.

How can we speed up this process quantumly? This is the subject of the
following sections.

\section{Quantum subroutines}

To speed up computations with one subroutine whose parameters are different in
the different cases we should run this subroutine for all these cases
simultaneously. But when trying to do this we meet the evident difficulty.
Quantum algorithm must transform the superposition $\sum\limits_i \la_i e_i$ 
of the basic states $e_i$. If a subroutine contains intermediate measurements
which require the essential classical action we must run this subroutine 
on each $e_i$ separatly which results in growth of time complexity as 
in classical case. Thus we need at least to exclude measurements from the
subroutine. But this is not sufficient. To use quantum parallelizm
for acceleration computation we must in addition to obtain the
particular form of output for our new subroutine. 
\n

{\bf Notation}
{\it For a real positive number $\e$
$\xi _\e $ denotes such state $\chi$ in Hilbert space that $\| \chi -\xi\| <\e$.}
\n

{\bf Definition}
{\it Unitary algorithm computing a function
$f:\ \{ 0,1\}^n \ar\{ 0,1\}$ with error probability $p_{err}$ is such quantum 
algorithm whose action on input data $x\in\{ 0,1\}^n$ is the sequence of
unitary transformations of the form
$$
\xi_0 \ar\xi_1 \ar\ldots\ar\xi_T ,
$$
where $\xi_0 =|x,0\rangle,\ \xi_T =(\tilde\xi \bigotimes|f(x)\rangle )_\e ,$
where $\e <p_{err} /2 ,\ \ \e$ depends on $x$.}
\nn

Thus the measurement of final state $\xi_T$ gives $f(x)$ with probability
greater than $1-p_{err}$. A unitary algorithm may be used as a subroutine because
the state $\sum\limits_i \la_i |x_i ,0\rangle$ is transformed to 
$\sum\limits_i \la_i\tilde\xi_i
\bigotimes f(x_i )\rangle +\bar\e ,$
where $\bar\e =\sum\limits_i \la_i \bar \e _{x_i}$ and $\|\bar e \| <p\sqrt{N} /2$,
where $N $ is the cardinality of all basic states. We assume that the time 
instant $T$ for the end of unitary algorithm is calculated classically beforehand
(look in the work \cite{De} of D. Deutsch ).

\section{Unitary quantum search}

Our nearest aim is to construct unitary algorithm with the time 
complexity $O(\sqrt{N_1 })$, for the standard problem of finding such $x$ that
$p(x)$ is true, for a given predicate $p, \ N_1 =2^{|x|}$.

\newtheorem{Proposition}{Proposition}
\begin{Proposition}
There is unitary algorithm which realizes the passage from
$|0,0,\ldots ,0\rangle$ to $|0,0,\ldots ,0,\g\rangle_\e$ with oracle $p$
where $\g = 1$ if $\exists x \ p(x)$ and $0$ else. This algorithm 
takes $\sqrt{N_1}$ time step with $Mn$ evaluations of $p$ at a time
and uses $M(n+2)+2$ qubits where
$M=\log (1/\e )$.
\end{Proposition}

    Proof

Recollect the algorithm G-BBHT (look at \cite{BBHT}).
It consists of the following steps.

1. The choise of number $m$.

2. The choise of value for integer variable $\chi$ distributed uniformly 
on $\{ 1,2,\ldots ,m\}$.

3. Perform Grover's transform $WF_0 WF_p$ $\ \ \chi$ times on initial state
$\sum\limits_j e_j /\sqrt{N}$.

4. Observe the result.

For our aim it is sufficient to take $m=\sqrt{N_1 }$.
With this value Lemma 2  from the work \cite{BBHT} says that
a required value $x$ will be the result of final observation with probability 
approximately $1/2$.
We need the following technical Lemma.
\newtheorem{Lemma}{Lemma}
\begin{Lemma}
For every $m=1,2,\ldots$ there exists unitary algorithm performing the passage:
$$
|0\ldots 0\rangle\ar\ldots\ar\frac{1}{\sqrt m}\sum\limits_
{\chi\leq m} |\chi\rangle ,
$$
where $\chi $ is an integer in its binary notation.
\end{Lemma}

This Lemma can be found in the work \cite{Ki} of A. Kitaev.

With Lemma 1 we can perform the points 1-3 of G-BBHT by unitary algorithm.

Now arrange $M$ independent blocks with $2n$ qubits each and fulfill
the unitary version of G-BBHT in each block independently. We obtain the state
$|0\ldots 0\rangle\bigotimes |x_1 x_2 \ldots x_M \rangle$
where the result
in every block is
\begin{equation}
x_i =\sum\limits_j \la_j e_j ,\ \
 \sum\limits_{p(e_j )\ \mbox{true}}\ |\la_j |^2
\approx 1/2 ,
\label{2}
\end{equation}
if such $e_j$ exists.
Applying oracle for $p$ we obtain the state
\begin{equation}
|0\ldots 0\rangle\bigotimes (\sum\limits_j
(\la_j |e_j \rangle\bigotimes | p(e_j )\rangle ))\bigotimes\ldots\bigotimes (\sum\limits_j
(\la_j |e_j \rangle\bigotimes |p(e_j )\rangle )) ,
\label{3}
\end{equation}
where $p(e_j )\in\{ 0,1\}$ are the values of ancillary qubits. Here we assume
that the oracle $p$ transforms $|a,b\rangle$ to $|a,b+p(a)\ \mbox{ (mod 2)}\rangle$. 

\begin{Lemma}{Lemma}
There exists a unitary algorithm with linear time 
complexity which fulfills the passage
$|\s_1 \s_2 \ldots \s_M 0^{M+2} \rangle \ar |\s_1
\s_2 \ldots \s_M 0^{M+1}\s\rangle$,
where $\forall i=1,2,\ldots ,M\ \ \s ,\s_i \in\{ 0,1\}$ and $\s =1$ iff
$\exists i\in\{ 1,\ldots ,M\} : \ \s_i =1$.
\end{Lemma}
\n

Proof of Lemma 2
\n

Consider a classical reversible transformation $f$ with three qubits:
\newline $|$ result, controller, subject $\rangle$, such that
$$
\begin{array}{l}
|\ 0\ 0\ 0\ \rangle \ \stackrel{f} \ar \ |\ 0\ 0\ 0\ \rangle\\
|\ 0\ 0\ 1\ \rangle \ \stackrel{f} \ar\ |\ 1\ 0\ 1\ \rangle\\
|\ 1\ 0\ 1\ \rangle \ \stackrel{f} \ar \ |\ 1\ 1\ 1\ \rangle\\
|\ 1\ 0\ 0\ \rangle \ \stackrel{f} \ar\ |\ 1\ 0\ 0\ \rangle \\
\end{array}
$$
Such transformation is unitary. If we apply it sequentially so that
on step number $ i $ the qubit "result" is always last ancillary one,
 "controller" is $ i $ -th ancillary qubit, "subject" is
 $\s _i$ and
$i=1,2,\ldots ,M,$ we obtain a state $|\d_1 \d_2 \ldots \d_M \d_{M+1}
\ldots\d_{2M} \s 0\rangle$. Then make $|\d_1 \d_2 \ldots \d_{2M} \s\s\rangle$ 
by $ |\s0\rangle \ar |\s\s\rangle ,$
and at last perform all reversal sequential transformations 
in reversal order which result in $|\s_1 \s_2 \ldots \s_M 0^{M+1} \s\rangle$.
Lemma 2 is proved.

We call the unitary algorithm from Lemma 2 EXISTS.
Now apply Lemma 2 to the state (3) with ancillary qubits playing 
the role of
$\s_1 ,\s_2 ,\ldots ,\s_M$. By (2) this results in the state 
\begin{equation}
(|0\ldots |0\rangle \bigotimes 
(\sum\limits_j \la_j |e_j \rangle\bigotimes |p(e_j )\rangle ))\bigotimes\ldots\bigotimes 
(\sum\limits_j (\la_j |e_j \rangle\bigotimes |p(e_j )\rangle ))\bigotimes\g )_\e .
\label{4}
\end{equation}
It is because if we have $M$ independent blocks of $n$ qubits each 
and perform our unitary algorithm on all these blocks independently we 
obtain a required value of $x$  at least in one block with probability
of order $1-\frac{1}{2^M}$, hence, having a value of admissible error $\e$,
$M=\log \frac{1}{\e}$ would suffice.
Now apply to all qubits but $\g$ in (4) reverse transformation to G-BBHT,
we obtain $(|0\ldots 0\g\rangle )_\e$.

Proposition 1 is proved.

{\bf Notation} {\it Denote the unitary algorithm from Proposition 1 with oracle $p$
by} SEARCH($p$).

Note that we can realize G-BBHT as unitary if $p$ is a given
unitary subroutine. 

\section{Formulas of predicate calculus}

Now take up a formula of predicate calculus of the form (1).
The generalization of it is a formula with free variables $z_1 ,z_2 ,\ldots ,
z_q$ of the form
\begin{equation}
\forall x_1 \exists x_2 \forall x_3 \ldots Q_{k-1} x_{k-1} Q_k x_k
p(z_1 , z_2 ,\ldots ,z_q ,x_1 ,\ldots ,x_k ).
\label{5}
\end{equation}
where $Q_1 ,Q_2 \in\{ \exists ,\forall\}$.
\n
\begin{Proposition}{Proposition}
There exists unitary algorithm which fulfills the passage
$$
|\ z_1 \ldots z_q  0\ldots 0\rangle \ar\ldots\ar | \ z_1 \ldots z_q
0\ldots 0 \g\rangle 
$$
in $ 2^{\frac{1}{2}\sum\limits_{i=1}^{k} |x_k |}$ steps using
$(Mn)^k$ queries at a time, where for every values of free variables 
$z_1 ,\ldots z_q$ 
$$
\g =\left\{
\begin{array}{l}
0,\ \mbox{if} \ (5) \ \mbox{true} ,\\
1,\ \mbox{if} \ (5) \ \mbox{false}.
\end{array}
\right.
$$
\end{Proposition}
\n

Proof
\n

Induction on $k$.
Basis. $k=0$. Nothing to prove. Step. Suppose it is true for the values of $k$
less than the given one, prove it for $k$. The inductive hypothesis says
that there exists a unitary algorithm with $ 2^{\frac{1}{2} 
\sum\limits_{i=1}^{k-1} |x_i | }$time complexity, no more than $(Mn)^{k-1}$ evaluations 
of $p$ at a time and $(Mn)^{k-1}$ 
qubits which computes the function 
\newline
$z_1  ,\ldots z_q , x_k \ar T_1 \in\{ 0,1\}$,
where $T_1 =1$ iff $\forall x_1 \exists x_2 \ldots  Q_{k-1} x_{k-1} p(z_1 ,\ldots ,z_q ,
x_1 , \ldots ,x_{k-1} ,x_k )$. Denote
this algorithm by P$_{k-1}$. Our aim is to construct the unitary algorithm 
for the function $z_1 ,\ldots ,z_q \ar T\in\{ 0,1\}$ , where $T=1$ iff

 $\forall x_1 \exists x_2 \ldots Q_{k-1} x_{k-1} Q_k x_k p(z_1 ,\ldots ,
z_q ,x_1 ,\ldots ,x_k )$.

Consider the different cases.

{\bf Case 1}: $Q_k$ is $\exists$.

Then the required algorithm is SEARCH (P$_{k-1}$). By Proposition 1 this
algorithm requires $2^{\frac{1}{2} |x_k |}$ time steps with $Mn$ simultaneous
evaluations of P$_{k-1}$
each of which by inductive hypothesis contains $ 2^{\frac{1}{2}
\sum\limits_{i=1}^{k-1} |x_i |}$ time steps with $(Mn)^{k-1}$
simultaneous evaluations of $p$. Hence we have a required
unitary algorithm with $ 2^{\frac{1}{2} \sum\limits_{i=1}^k |x_1|}$ time steps
and $(Mn)^k$ simultaneous
evaluations of $p$. The number of required qubits will be of order $(Mn)^k$.
\n

{\bf Case 2}: $Q_k$ is $\forall$.

Then the required algorithm is NOT (SEARCH (NOT (P$_{k-1}$ ))), where 
NOT is the negation $0\ar 1 ,\ 1\ar 0$.
Proposition 2 is proved.

Theorem is a particular case of Proposition 2. Theorem is proved.

\section{Conclusion}

We see that if a formula of predicate calculus has a limited number 
$k$ of quantor changes then the quantum verification of it requires the time
of order $\sqrt{N}$ and polynomial number of qubits, 
where $N$ is the time required for classical 
verifivation.

The open question is: can this fact be generalized to the case of 
arbitrary finite number of quantor changes, e.g. when $k\ar\infty$, or not.

\section{Acknowledgements}

I am grateful to Lov Grover for the useful discussion and attention to my work
and to the principal of "Stankin" Yuri Solomentsev for the financial support
of my work. The reference to the work 9802040 was absent 
in the first version of this work and I am grateful to Richard Cleve and 
Harry Buhrman who informed me about it.

\end{document}